\documentclass[11pt]{amsart}
\usepackage{amssymb}

\theoremstyle{plain}
\newtheorem{thm}{Theorem}[section]
\newtheorem{lem}[thm]{Lemma}
\theoremstyle{definition}
\newtheorem{conj}{Conjecture}
\newtheorem{claim}{Claim}[thm]
\newtheorem{subclaim}{Subclaim}[claim]
\newtheorem{case}{Case}[thm]
\newtheorem{subcase}{Subcase}[case]
\newtheorem{subsubcase}{Subsubcase}[subcase]

\renewcommand{\theclaim}{\arabic{claim}}
\renewcommand{\thesubclaim}{\arabic{claim}.\arabic{subclaim}}

\newcommand{\qedof}[1]%
{\hspace*{\fill}\mbox{$\Box$(#1 \number\thethm)}\newline}
\newenvironment{proofofthm}{{\it Proof.}}{\qedof{Theorem}}
\newenvironment{proofoflem}{{\it Proof.}}{\qedof{Lemma}}

\newenvironment{proofof*}[1]{{{\it Proof of #1.}}\def\proving{#1}}%
{\hspace*{\fill}\mbox{$\Box$(\proving)}\newline}
\newenvironment{proofofclaim}{\raisebox{-.4ex}{\Large $\vdash$\ \ }}%
{\mbox{}\hspace*{\fill}\raisebox{-.4ex}{\Large $\dashv$ }\nolinebreak%
\mbox{(Claim \number\theclaim)}\newline}
\newenvironment{proofofsubclaim}{\raisebox{-.4ex}{\Large $\vdash$\ \ }}%
{\mbox{}\hspace*{\fill}\raisebox{-.4ex}{\Large $\dashv$ }\nolinebreak%
\mbox{(Subclaim \number\thesubclaim)}\newline}

\newcommand{\nin}{\not\in}

\newcommand{\N}{\mathbb{N}}
\newcommand{\Z}{\mathbb{Z}}

\newcommand{\Seq}[1]{\langle #1\rangle}

\newcommand{\vecx}{\vec{x}}
\newcommand{\vecy}{\vec{y}}
\newcommand{\vecz}{\vec{z}}

\newcommand{\btext}[1]{{\text{\bf{#1}}}}
\newcommand{\bif}{\btext{ if }}
\newcommand{\bthen}{\btext{ then }}
\newcommand{\belse}{\btext{ else }}
\newcommand{\belseif}{\btext{ else if }}

\begin{document}
\title{Notes on higher-dimensional tarai functions}
\author{Tetsuya Ishiu}
\email{ishiut@muohio.edu}
\address{Department of Mathematics, 
  Miami University, 
  Oxford, OH, 45056, USA}
\thanks{This material is based upon work supported 
  by the National Science Foundation under Grant No.~0700983.}

\author{Masashi Kikuchi}
\email{kikuchi@lepidum.co.jp}
\address{1-28-7 Sasazuka, Shibuya-ku, Tokyo, 151-0073, Japan}
\date{\today}

\maketitle
\setcounter{section}{-1}
\section{Introduction}

I.~Takeuchi defined the following recursive function, 
called the tarai function, in
\cite{takeuchi:_recur_funct_that_does_almos_recur_only}. 
\begin{align*}
 t(x, y, z)&=\bif x\leq y\bthen y
 \belse t(t(x-1, y, z), t(y-1, z, x), t(z-1, x, y)) 
\end{align*}
This function requires many recursive calls even for 
small $x$, $y$, and $z$, so it is used to see how 
effectively the programming language implementation 
handles recursive calls. 
In \cite{mccarthy78:_inter_lisp_funct}, 
J. McCarthy proved that this recursion terminates 
without call-by-need and 
$t$ can be computed in the following way. 
\begin{align*}
 t(x, y, z)&=\bif x\leq y\bthen y\belseif y\leq z\bthen z\belse x
\end{align*}
In \cite{moore79:_mechan_proof_of_termin_of_takeuc_funct}, 
J.~S. Moore gave an easier proof and verified it 
by the Boyer-Moore theorem prover. 

D. Knuth proposed the following generalization in 
\cite{knuth91:_textb_examp_of_recur}, called 
the $n$-dimensional tarai function. 
\begin{align*}
 t(x_1, x_2, \ldots, x_n)&=
 \bif x_1\leq x_2 \bthen x_2 \\
 &\:\:\belse
 t(t(x_1-1, x_2, \ldots, x_n), \ldots, 
  t(x_n-1, x_1, \ldots, x_{n-1}))
\end{align*}
It was shown by T.~Bailey, J.~Coldwell, and J.~Cowles 
that the $4$-dimensional tarai function does not terminate 
without call-by-need because 
\begin{align*}
  t(3, 2, 1, 5)&=
  t(t(2, 2, 1, 5), t(1, 1, 5, 3), t(0, 5, 3, 2), t(4, 3, 2, 1))\\
  &=t(2, 1, 5, 4)\\
  &=t(t(1, 1, 5, 4), t(0, 5, 4, 2), t(4, 4, 2, 1), t(3, 2, 1, 5))
\end{align*}

T. Bailey and J. Cowles announced in \cite{Bailey_Cowles_tarai} 
that they gave an informal (handwritten) proof 
of the following conjecture. 

\begin{conj}
  Let $n\geq 3$ be an integer. 
  Define the function $f$ on $\Z^n$  by 
  \begin{align*}
    f(x_1, x_2, \ldots, x_m)&=\bif (\exists k<m)(x_1>x_2>\cdots>x_k\leq x_{k+1})\\
    &\phantom{\bif} \bthen g_b(x_1, x_2, \ldots, x_{k+1})\\
    &\phantom{\bif} \belse x_1
  \end{align*}
  where the function $g_b$ is defined by 
  \begin{align*}
    g_b(x_1, x_2, \ldots, x_j)&=
    \bif j\leq 3\bthen x_j\\
    &\phantom{=\bif j\leq 3}\belse \bif x_1=x_2+1\text{ or }x_2>x_3+1\\
    &\phantom{=\bif j\leq 3\belse \bif }\bthen g_b(x_2, \ldots, x_j)\\
    &\phantom{=\bif j\leq 3}\belse \max\{x_3, x_j\}
  \end{align*}
  Then, $f$ satisfies the $n$-dimensional tarai recurrence. 
\end{conj}

The goal of this paper is to give a proof to this theorem. 
Moreover, the proof will be simpler than the one proposed 
in \cite{Bailey_Cowles_tarai}, so we hope that it is easier 
to be formalized. 

\section{Termination with call-by-need}
\label{sec:term-with-call}

In this section, we shall prove that 
the $n$-dimensional tarai function is a total function 
for every $n\geq 3$. 
Throughout this section, let 
$n$ be a fixed natural number with $n\geq 3$ and 
$t$ the $n$-dimensional tarai function. 

First, we shall prepare notation. 
Let $\vecx=\Seq{x_1, \cdots, x_n}\in\Z^n$. 
Define $\sigma(\vecx)$ and $r(\vecx)$ by

\begin{align*}
 \sigma(\vecx)&=\Seq{x_1-1, x_2, \cdots, x_n}\\
 r(\vecx)&=\Seq{x_2, x_3, \cdots, x_n, x_1}
\end{align*}

Namely, for every $i\in\{1, \cdots, n\}$, 
\begin{align*}
 \sigma(\vecx)(i)&=
 \begin{cases}
  \vecx(1)-1 & \text{ if }i=1\\
  \vecx(i) & \text{ otherwise}
 \end{cases}
\end{align*}
and 
\begin{align*}
 r(\vecx)(i)&=
 \begin{cases}
  \vecx(i+1) & \text{ if }i<n\\
  \vecx(1) & \text{ if }i=n
 \end{cases}
\end{align*}
In particular, for every $i\in\{1, \cdots, n-1\}$ 
and $j\in\{1, \cdots, n\}$
\begin{align*}
 r^i(\vecx)(j) &=
 \begin{cases}
  \vecx(j+i) &\text{ if }j+i\leq n\\
  \vecx(j+i-n) & \text{ if }j+i>n
 \end{cases}
\end{align*}

By using this notation, 
the $n$-dimensional tarai function $t$ can be 
defined as for every $\vec{x}\in\Z^n$, 
if $\vec{x}(1)\leq\vec{x}(2)$, then $t(\vec{x})=\vec{x}(2)$, and 
if $\vec{x}(1)>\vec{x}(2)$, then 
$t(\vec{x})=t(\vec{y})$, where 
$\vec{y}\in\Z^n$ is defined by 
$\vec{y}(i)=t(\sigma(r^{i-1}(\vec{x})))$ 
for all $i=1, 2, \ldots, n$. 

Let $k\in\N$ with $2\leq k\leq n$. 
Define $X_k$ to be the set of all 
$\vecx\in\Z^n$ such that 
for every $i<k$, $\vecx(i)\leq\vecx(k)$. 
Note that the family $\{X_k : 2\leq k\leq n\}$ 
is not pairwise disjoint. 
For example, $\Seq{2, 1, 4, 3, 5}\in X_3\cap X_5$. 
Note also that if $\vec{x}\in\Z^n$ satisfies 
$\vec{x}(1)<\max\vec{x}$, 
then there exists a $k$ such that 
$\vec{x}\in X_k$. 

\begin{lem}
 \label{lem:X_k}
 For every $k\in\N$ with $2\leq k\leq n$, 
 the following statement holds. 
 \begin{itemize}
  \item[$(*)_k$:] For every $\vecx\in X_k$, 
		  \begin{enumerate}
		   \item $t(\vecx)$ terminates with call-by-need,   
		   \item $t(\vecx)$ depends only on 
			 $\vecx\restriction\{1, \cdots, k\}$, and 
		   \item $t(\vecx)\leq\vecx(k)$. 
		  \end{enumerate}
 \end{itemize}
\end{lem}

\begin{proofoflem}
 Go by induction on $k$. 
 
 First assume $k=2$. 
 Then, we have $\vecx(1)\leq\vecx(2)$ and hence 
 $t(\vecx)=\vecx(2)$. 
 Hence, $\vecx$ clearly satisfies (i)--(iii). 

 Suppose that $(*)_{k'}$ holds for all $k'\in\N$ with $2\leq k'<k$. 
 We shall prove $(*)_k$. 
 By way of contradiction, 
 suppose that there exists an $\vecx\in X_k$ which does not 
 satisfy one of (i)--(iii). 
 By inductive hypothesis, we have $\vecx\nin X_{k-1}$. 
 In particular, $\vecx(1)>\vecx(2)$. 
 Hence, we can pick the least $y_1\in\Z$ such that 
 if $\vecy$ is defined as 
 $\vecy(1)=y_1$ and $\vecy(i)=\vecx(i)$ for every 
 $i\in\{2, \cdots, n\}$, 
 then $\vecy$ does not satisfy one of (i)--(iii). 
 By redefining $\vecx$, we may assume that 
 for every $\vecx'\in\N^n$, if 
 $\vecx'(1)<\vecx(1)$ and $\vecx'(i)=\vecx(i)$ for every 
 $i\in\{2, \cdots, n\}$, 
 then $\vecx'$ satisfies (i)--(iii). 

 For each $i\in\{1, \ldots, k\}$, 
 define $\vecz_i=\sigma(r^{i-1}(\vecy))$ and 
 $\vecy(i)=t(\vecz_i)$. 
 For $i\in\{k+1, \ldots, n\}$, 
 let $\vecy(i)$ be left undefined. 
 By definition, 
 $t(\vecx)$ terminates with call-by-need if 
 $t(\vecz_i)$ for all $i\in\{1, \ldots, k\}$ and 
 $t(\vecy)$ terminate with call-by-need, and 
 in that case, $t(\vecx)=t(\vecy)$. 

 Notice that for every $i\in\{1, \cdots, k-1\}$ 
 and $j\in\{2, \cdots, k-i+1\}$, 
 \begin{align*}
  \vecz_i(j)&=\vecx(j+i-1)\leq\vecx(k)\\
  \vecz_i(1)&=\vecx(i)-1\leq\vecx(k)\\
  \intertext{and}
  \vecz_i(k-i+1)&=\vecx(i+k-i+1-1)=\vecx(k)
 \end{align*}
 Hence, 
 $\vecz_i(k-i+1)=\vecx(k)$. 
 Therefore, $\vecz_i\in X_{k-i+1}$. 
 If $i\geq 2$, then by inductive hypothesis, 
 $(*)_{k-i+1}$ holds. 
 Thus, $\vecz_i$ satisfies (i)--(iii).
 In particular, 
 $t(\vecz_i)\leq\vecz_i(k-i+1)=\vecx(k)$. 
 If $i=1$, we have $\vecz_1=\sigma(\vecx)$ and 
 by the minimality of $\vecx(1)$, we know that 
 $\vecz_1$ satisfies (i)--(iii). 
 For every $i\in\{1, \cdots, k-1\}$, 
 $\vecy(i)=t(\vecz_i)\leq\vecx(k)$. 
 Moreover, 
 $\vecz_{k-1}(2)=\vecx(k)$. 
 Thus, $\vecz_{k-1}(1)\leq\vecx(k)=\vecz_{k-1}(2)$, 
 which implies 
 $\vecy(k-1)=t(\vecz_{k-1})=\vecz_{k-1}(2)=\vecx(k)$. 
 Hence, $\vecy\in X_{k-1}$ and so 
 $\vecy$ satisfies (i)--(iii). 
 It is now easy to see that $\vecx$ also satisfies (i)--(iii). 
\end{proofoflem}

\begin{thm}
 \label{thm:termination}
 For every $\vecx\in\Z^n$, $t(\vecx)$ terminates with call-by-need 
 and $t(\vecx)\leq\max(\vecx)$. 
\end{thm}

\begin{proofofthm}
 By Lemma \ref{lem:X_k}, 
 it suffices to show that for every $\vecx\in\Z^n$ with 
 $\vecx(1)=\max(\vecx)$, 
 $t(\vecx)$ terminates with call-by-need.  

 By way of contradiction, 
 suppose that $t(\vecx)$ does not terminate. 
 We may also assume that 
 for every $\vecx'\in\Z^n$ with 
 $\vecx'(1)<\vecx(1)$ and 
 $\vecx'(i)=\vecx(i)$ for every $i\in\{2, \cdots, n\}$, 
 $t(\vecx')$ terminates with call-by-need. 

 For every $i\in\{1, \cdots, n\}$, 
 let $\vecz_i=\sigma(r^{i-1}(\vecx))$ and 
 $\vecy(i)=t(\vecz_i)$. 
 First suppose $i\geq 2$. 
 Then, for every $j\in\{2, \cdots, n-i+1\}$, 
 \begin{align*}
  \vecz_i(j)&=\vecx(j+i-1)\leq\vecx(1)
 \end{align*}
 and 
 \begin{align*}
  \vecz_i(1)=\vecx(i)-1\leq\vecx(1)
 \end{align*}
 Moreover, 
 $\vecz_i(n-i+2)=\vecx(n-i+2+(i-1)-n)
 =\vecx(1)$. 
 Thus, $\vecz_i\in X_{n-i+2}$. 
 Hence, $t(\vecz_i)$ terminates with call-by-need and 
 $t(\vecz_i)\leq\vecx(1)$. 
 In addition, 
 $\vecz_n(1)=\vecx(1+(n-1))=\vecx(n)\leq\vecx(1)$ and 
 $\vecz_n(2)=\vecx(2+(n-1)-n)=\vecx(1)$. 
 Thus, $t(\vecz_n)=\vecx(1)$. 
 
 Suppose that $i=1$. 
 Then, $\vecz_1=\sigma(\vecx)$. 
 By the minimality of $\vecx(1)$, 
 $t(\vecz_1)$ terminates with call-by-need and $t(\vecz_1)\leq\vecx(1)$. 

 Therefore, for every $i\in\{1, \cdots, n\}$, 
 we have $\vecy(i)=t(\vecz_i)\leq\vecx(1)$ and $t(\vecz_n)=\vecx(1)$. 
 Hence, we have $\vecy\in X_n$. 
 By $(*)_n$, $t(\vecy)$ terminates with call-by-need and 
 $t(\vecy)\leq\vecy(n)=\vecx(1)$. 
 It follows that $t(\vecx)$ also terminates with call-by-need and 
 $t(\vecx)\leq\vecx(1)$. 
  This contradicts the choice of $\vecx$. 
\end{proofofthm}

\section{Alternative definition of the function of T. Bailey and J. Cowles}

In this section, we shall set up some definitions and notation 
which help us prove the main theorem. 
Let $F$ denote the set of all non-empty finite sequences of 
integers. 
$f$ denotes the function defined by T. Bailey and J. Cowles. 

Let $k$ be a function with domain $F$ as follows. 
Let $\vecx\in F$ be of length $n$. 
If $\vecx(1)>\vecx(2)>\ldots>\vecx(n)$, 
then let $k(\vecx)=n$. 
Otherwise, let $k(\vecx)$ be the least $k$ such that 
$\vecx(k)\leq\vecx(k+1)$. 

Let $l$ be a function with domain $F$ as follows. 
Let $\vecx\in F$ be of length $n$. 
If there is an integer $l$ with $1\leq l<k(\vecx)$ 
such that 
$\vecx(l)>\vecx(l+1)+1$ and $\vecx(l+1)=\vecx(l+2)+1$, 
then let $l(\vecx)$ be the least such $l$. 
Otherwise, let $l(\vecx)=k(\vecx)-1$. 
Notice that $l(\vecx)=0$ if and only if $k(\vecx)=1$. 

\begin{lem}
 \label{lem:g_b}
 Let $\vecx\in F$ be of length $n\geq 3$. 
 Suppose that $k(\vecx)=n-1$. 
 Then, 
 $g_b(\vecx)=\max(\vecx(l(\vecx)+2), \vecx(k(\vecx)+1))$. 
\end{lem}

\begin{proofoflem}
 We shall prove the lemma by induction on $n$. 
 If $n=3$, then $g_b(\vecx)=\vecx(3)$. 
 We also have $k(\vecx)=2$ and $l(\vecx)=1$. 
 Thus, $\max(\vecx(l(\vecx)+2), \vecx(k(\vecx)+1))=\vecx(3)$. 
 Therefore, $g_b(\vecx)=\max(\vecx(l(\vecx)+2), \vecx(k(\vecx)+1))$. 

 Suppose that the conclusion holds for all $\vecx$ of length $n$ 
 for some $n\geq 3$. 
 Let $\vecx\in F$ be of length $n+1$ with 
 $k(\vecx)=n$. 
 First suppose that $\vecx(1)>\vecx(2)+1$ and $\vecx(2)=\vecx(3)+1$. 
 Then $g_b(\vecx)=\max(\vecx(3), \vecx(n+1))$. 
 It is clear that $l(\vecx)=1$. Thus, 
 $\max(\vecx(l(\vecx)+2), \vecx(k(\vecx)+1))
 =\max(\vecx(3), \vecx(n+1))=g_b(\vecx)$. 
 
 Suppose $\vecx(1)=\vecx(2)+1$ or $\vecx(2)>\vecx(3)+1$. 
 Then $g_b(\vecx)=g_b(\vecy)$ where 
 $\vecy$ is a sequence of length $n$ such that 
 $\vecy(i)=\vecx(i+1)$ for every $i=1, \ldots, n$. 
 Notice that $k(\vecy)=k(\vecx)-1$ and 
 $l(\vecy)=l(\vecx)-1$. 
 By inductive hypothesis, 
 $g_b(\vecy)=\max(\vecy(l(\vecy)+2), \vecy(k(\vecy)+1))
 =\max(\vecy(l(\vecx)+1), \vecy(k(\vecx)))
 =\max(\vecx(l(\vecx)+2), \vecx(k(\vecx)+1))$. 
 Therefore, 
 $g_b(\vecx)=\max(\vecx(l(\vecx)+2), \vecx(k(\vecx)+1))$
\end{proofoflem}

By using the previous lemma, the following is immediate. 

\begin{lem}
 Let $\vecx\in F$ be of length $n\geq 3$. 
 If $k(\vecx)=n$ (i.e. $\vecx(1)>\vecx(2)>\ldots>\vecx(n)$), 
 then $f(\vecx)=\vecx(1)$. 
 If $k(\vecx)<n$, then 
 $f(\vecx)=\max(\vecx(l(\vecx)+2), \vecx(k(\vecx)+1))$. 
\end{lem}

We shall use this characterization in the next section. 

\section{Closed form}
\label{sec:r_closed_form}

In this section, we shall give a proof of the theorem of 
T. Bailey and J. Cowles. 
Note that if carefully rewritten, the proof simultaneously 
gives the termination of the $n$-dimensional tarai function. 
We chose to give separate proofs to simplify the arguments. 
Throughout this section, we fix a natural number $n$ 
with $n\geq 3$. 

We shall show that for every $\vecx\in\Z^n$, 
$t(\vecx)=f(\vecx)$. 
To this end, by Theorem \ref{thm:termination}, 
it suffices to show that $f$ satisfies the $n$-dimensional 
tarai recurrence. 

We shall begin with some easy facts about $f$. 

\begin{lem}
  \label{lem:I}
  For every $\vecx\in\Z^n$, 
  if $k(\vecx)\leq 2$, then 
  $f(\vecx)=\vecx(k(\vecx)+1)$. 
\end{lem}

\begin{proofoflem}
  Since $k(\vecx)\leq 2$, we have $l(\vecx)=k(\vecx)-1$. 
  Thus, $f(\vecx)=\max\{\vecx(l(\vecx)+2), \vecx(k(\vecx)+1)\}=\vecx(k(\vecx)+1)$. 
\end{proofoflem}

\begin{lem}
  \label{lem:A}
  For every $\vecx\in\Z^n$ and $m\in\{1, \ldots, l(\vecx)+2\}$, 
  if $k(\vecx)<n$, and 
  $\vecx(m)\leq\vecx(k(\vecx)+1)$, 
  then $f(\vecx)=\vecx(k(\vecx)+1)$. 
\end{lem}

\begin{proofoflem}
  If $l(\vecx)=k(\vecx)-1$, then clearly 
  $f(\vecx)=\max\{\vecx(l(\vecx)+2), \vecx(k(\vecx)+1)\}
  =\vecx(k(\vecx)+1)$. 
  Suppose $l(\vecx)<k(\vecx)-1$. 
  Then $l(\vecx)+2\leq k(\vecx)$. 
  We have $m\leq l(\vecx)+2\leq k(\vecx)$. 
  Thus, 
  \begin{align*}
    \vecx(l(\vecx)+2)&\leq\vecx(m)\leq\vecx(k(\vecx)+1)
  \end{align*}
  So, $f(\vecx)=\max\{\vecx(l(\vecx)+2), \vecx(k(\vecx)+1)\}=\vecx(k(\vecx)+1)$.   
\end{proofoflem}

\begin{lem}
  \label{lem:E}
  For every $\vecx\in\Z^n$, 
  if $k(\vecx)<n$ and $\vecx(3)\leq\vecx(k(\vecx)+1)$, 
  then $f(\vecx)=\vecx(k(\vecx)+1)$. 
\end{lem}

\begin{proofoflem}
  By Lemma \ref{lem:I}, if $k(\vecx)\leq 2$, then 
  $f(\vecx)=\vecx(k(\vecx)+1)$. 
  Suppose $k(\vecx)\geq 3$. 
  Then by applying Lemma \ref{lem:A} with $m=3$, 
  we have $f(\vecx)=\vecx(k(\vecx)+1)$. 
\end{proofoflem}

\begin{lem}
  \label{lem:H}
  For every $\vecx\in\Z^n$, 
  if $k(\vecx)<n$ and $\vecx(2)\leq\vecx(k(\vecx)+1)$, 
  then $f(\vecx)=\vecx(k(\vecx)+1)$. 
\end{lem}

\begin{proofoflem}
  If $k(\vecx)\geq 3$, then 
  we have $\vecx(3)<\vecx(2)\leq\vecx(k(\vecx)+1)$. 
  By Lemma \ref{lem:E}, 
  $f(\vecx)=\vecx(k(\vecx)+1)$. 
  If $k(\vecx)\leq 2$, then by Lemma \ref{lem:I}, 
  we have $f(\vecx)=\vecx(k(\vecx)+1)$. 
\end{proofoflem}

\begin{lem}
  \label{lem:B}
  For every $\vecx\in\Z^n$, 
  if $k(\vecx)<n$ and $l(\vecx)\geq k(\vecx)-2$, 
  then $f(\vecx)=\vecx(k(\vecx)+1)$. 
\end{lem}

\begin{proofoflem}
  By definition, $l(\vecx)\leq k(\vecx)-1$. 
  So, $l(\vecx)\geq k(\vecx)-2$ 
  implies either $l(\vecx)=k(\vecx)-2$ or $l(\vecx)=k(\vecx)-1$. 
  If $l(\vecx)=k(\vecx)-1$, 
  then clearly 
  $f(\vecx)=\vecx(k(\vecx)+1)$. 
  Suppose $l(\vecx)=k(\vecx)-2$. 
  Then, 
  \begin{align*}
    f(\vecx)&=\max\{\vecx(l(\vecx)+2), \vecx(k(\vecx)+1)\}\\
    &=\max\{\vecx(k(\vecx)), \vecx(k(\vecx)+1)\}\\
    &=\vecx(k(\vecx)+1)
  \end{align*}
  since by the definition of $k(\vecx)$, 
  $\vecx(k(\vecx))\leq\vecx(k(\vecx)+1)$. 
\end{proofoflem}

\begin{lem}
  \label{lem:C}
  For every $\vecx\in\Z^n$, if $\vecx(2)\leq\vecx(3)$, 
  then $f(\vecx)$ is either $\vecx(2)$ or $\vecx(3)$. 
  In particular, if $\vecx(2)=\vecx(3)$, then $f(\vecx)=\vecx(2)$. 
\end{lem}

\begin{proofoflem}
  Since $\vecx(2)\leq\vecx(3)$, we have $k(\vecx)\leq 2$. 
  Then, $k(\vecx)-2\leq 0\leq l(\vecx)$. 
  By Lemma \ref{lem:B}, $f(\vecx)=\vecx(k(\vecx)+1)$. 
  Since $k(\vecx)$ is either $1$ or $2$, 
  $f(\vecx)$ is either $\vecx(2)$ or $\vecx(3)$. 
\end{proofoflem}

\begin{lem}
  \label{lem:D}
  For every $\vecx\in\Z^n$ and $m\in\{1, \ldots, n-1\}$, 
  if $\vecx(m)=\vecx(m+1)$, $k(\vecx)\geq m-1$, and $l(\vecx)\geq m-2$, 
  then $f(\vecx)=\vecx(m)$. 
\end{lem}

\begin{proofoflem}
  Since $\vecx(m)=\vecx(m+1)$, we have $k(\vecx)\leq m$. 
  So, 
  \begin{align*}
    l(\vecx)&\geq m-2\geq k(\vecx)-2
  \end{align*}
  Hence, by Lemma \ref{lem:B}, $f(\vecx)=\vecx(k(\vecx)+1)$. 
  However, since $m-1\leq k(\vecx)\leq m$, 
  $k(\vecx)$ is either $m-1$ or $m$. 
  If $k(\vecx)=m-1$, then $\vecx(k(\vecx)+1)=\vecx(m)$. 
  If $k(\vecx)=m$, then $\vecx(k(\vecx)+1)=\vecx(m+1)=\vecx(m)$ 
  by assumption. 
\end{proofoflem}

\begin{lem}
  \label{lem:F}
  For every $\vecx\in\Z^n$ and $m\in\{1, \ldots, n-2\}$, 
  if $k(\vecx)\geq m-1$, $l(\vecx)\geq m-2$,  
  $\vecx(m)=\vecx(m+2)$, and 
  $\vecx(m)\geq\vecx(m+1)$, 
  then $f(\vecx)=\vecx(m)$. 
\end{lem}

\begin{proofoflem}
  By assumption, $\vecx(m+1)\leq\vecx(m)=\vecx(m+2)$. 
  So, $k(\vecx)\leq m+1$. 
  Therefore, $k(\vecx)$ is either $m-1$, $m$, or $m+1$. 
  \setcounter{case}{0}
  \begin{case}
    $k(\vecx)=m-1$. 
  \end{case}
  Then, $l(\vecx)\geq m-2=k(\vecx)-1$. 
  By Lemma \ref{lem:B}, 
  $f(\vecx)=\vecx(k(\vecx)+1)=\vecx(m)$. 

  \begin{case}
    $k(\vecx)=m$. 
  \end{case}
  In this case, we have $\vecx(m)\leq \vecx(m+1)$. 
  By assumption, we also have $\vecx(m+1)\leq\vecx(m)$. 
  Therefore, $\vecx(m)=\vecx(m+1)$. 
  By Lemma \ref{lem:D}, 
  $f(\vecx)=\vecx(m)$. 

  \begin{case}
    $k(\vecx)=m+1$ and $l(\vecx)=m-2$. 
  \end{case}

  Then, $\vecx(k(\vecx)+1)=\vecx(m+2)=\vecx(m)$ 
  and $\vecx(l(\vecx)+2)=\vecx(m)$. 
  Thus, $f(\vecx)=\vecx(m)$. 

  \begin{case}
    $k(\vecx)=m+1$ and $l(\vecx)\geq m-1$. 
  \end{case}

  Note that $l(\vecx)\geq m-1=k(\vecx)-2$. 
  By Lemma \ref{lem:B}, $f(\vecx)=\vecx(k(\vecx)+1)=\vecx(m+2)=\vecx(m)$. 
\end{proofoflem}

\begin{lem}
  \label{lem:G}
  For every $\vecx\in\Z^n$, if $2\leq k(\vecx)<n$, 
  then $f(\vecx)\leq\max\{\vecx(3), \vecx(k(\vecx)+1)\}$. 
\end{lem}

\begin{proofoflem}
  If $f(\vecx)=\vecx(k(\vecx)+1)$, it is trivial. 
  Thus, we assume that $f(\vecx)=\vecx(l(\vecx)+2)\neq\vecx(k(\vecx)+1)$. 
  It is easy to see that $l(\vecx)\leq k(\vecx)-2$. 
  So, $l(\vecx)+2\leq k(\vecx)$. 
  If $l(\vecx)=0$, then $k(\vecx)=1$, which contradicts $l(\vecx)\leq k(\vecx)-2$. 
  So, we have $l(\vecx)\geq 1$. 
  Therefore, $3\leq l(\vecx)+2\leq k(\vecx)$. 
  Hence, $\vecx(3)\geq\vecx(l(\vecx)+2)=f(\vecx)$. 
\end{proofoflem}



\begin{lem}
  \label{k(x)_lt_n}
  For every $\vecx\in\Z^n$, if $k(\vecx)<n$, then 
  $f(\vecx)$ satisfies the tarai recurrence. 
\end{lem}

\begin{proofoflem}
  Let $k=k(\vecx)$ and $l=l(\vecx)$. 
  If $k=1$, it is trivial. Assume $k\geq 2$. 

 For each $i=1, \cdots, n$, 
 define $\vecz_i=\sigma(r^{i-1}(\vecx))$ and 
 let $\vecy\in\Z^n$ be defined by
 $\vecy(i)=f(\vecz_i)$. 
 It suffices to show 
 $f(\vecx)=f(\vecy)$. 

 It is easy to see that 
 \begin{align*}
  \vecz_i(j)&=
  \begin{cases}
   \vecx(i)-1 &\text{ if }j=1\\
   \vecx(j+i-1) & \text{ if }j+i-1\leq n\\
   \vecx(j+i-1-n)& \text{ if }j+i-1>n
  \end{cases}
 \end{align*}

 Define $m$ to be the least such that 
 $\vecx(m)>\vecx(m+1)+1$ or $m=k-1$. 
 Clearly we have $m\leq l$. 

  \begin{claim}
    \label{claim:1}
    $\vecy(k)=\vecx(k+1)$
  \end{claim}

  \begin{proofofclaim}
    Note that 
    $\vecz_k(1)=\vecx(k)-1$ and $\vecz_k(2)=\vecx(k+1)$. 
    Since $\vecx(k)\leq\vecx(k+1)$, we have
    $\vecz_k(1)\leq\vecz_k(2)$. 
    Hence, $\vecy(k)=f(\vecz_k)=\vecz_k(2)=\vecx(k+1)$. 
  \end{proofofclaim}

  \begin{claim}
    \label{claim:2}
    $k(\vecy)\leq k-1$. 
  \end{claim}

  \begin{proofofclaim}
    Note
    $\vecz_{k-1}(2)=\vecx(k)$ 
    and $\vecz_{k-1}(3)=\vecx(k+1)$. 
    So, $\vecz_{k-1}(2)\leq\vecz_{k-1}(3)$. 
    By Lemma \ref{lem:C}, 
    $\vecy(k-1)$ is either $\vecz_{k-1}(2)=\vecx(k)$ 
    or $\vecz_{k-1}(3)=\vecx(k+1)$. 
    In either way, we have $\vecy(k-1)\leq\vecx(k+1)=\vecy(k)$. 
    Thus, $k(\vecy)\leq k-1$. 
  \end{proofofclaim}

  \begin{claim}
    \label{claim:3}
    For every $i\in\{1, \ldots, n-1\}$, 
    if $\vecx(i)=\vecx(i+1)+1$, then $\vecy(i)=\vecx(i+1)$. 
    In particular, if $i<m$, then $\vecy(i)=\vecx(i+1)$. 
  \end{claim}

  \begin{proofofclaim}
    We have $\vecz_i(1)=\vecx(i)-1$ and 
    $\vecz_i(2)=\vecx(i+1)$. 
    Thus, $\vecz_i(1)=\vecz_i(2)$. 
    So, 
    $\vecy(i)=\vecz_i(2)=\vecx(i+1)$. 
  \end{proofofclaim}

  \begin{claim}
    \label{claim:4}
    For every $i\in\{1, \ldots, m-2\}$, 
    $\vecy(i)=\vecy(i+1)+1$. 
    In particular, $k(\vecy)\geq m-1$ and $l(\vecy)\geq m-2$. 
    If $l(\vecy)<k(\vecy)-1$, then $l(\vecy)\geq m-1$. 
  \end{claim}

  \begin{proofofclaim}
    If $i\in\{1, \ldots, m-2\}$, then 
    by Claim \ref{claim:3}, 
    $\vecy(i)=\vecx(i+1)$ and $\vecy(i+1)=\vecx(i+2)$. 
    Since $i+1<m$, we have $\vecx(i+1)=\vecx(i+2)+1$. 
    Thus, $\vecy(i)=\vecy(i+1)+1$. 
    So we have $k(\vecy)\geq m-1$. 
    If $l(\vecy)=k(\vecy)-1$, then $l(\vecy)\geq m-1-1=m-2$. 
    If $l(\vecy)<k(\vecy)-1$, then 
    we have $\vecy(l(\vecy))>\vecy(l(\vecy)+1)+1$. 
    But we also have $\vecy(m-2)=\vecy(m-1)$ 
    and hence $l(\vecy)\neq m-2$. 
    So, $l(\vecy)\geq m-1$. 
  \end{proofofclaim}

  \begin{claim}
    \label{claim:5}
    For every $i\in\{1, \ldots, k-1\}$, 
    if $\vecx(i)>\vecx(i+1)+1$, 
    then $k(\vecz_i)=k-i+1$ and 
    $\vecz_i(k(\vecz_i)+1)=\vecx(k+1)$. 
  \end{claim}

  \begin{proofofclaim}
    Note $\vecz_i(1)=\vecx(i)-1$ and $\vecz_i(2)=\vecx(i+1)$. 
    By assumption, $\vecz_i(1)>\vecz_i(2)$. 
    For every $j\in\{2, \ldots, k-i\}$, 
    we have $\vecz_i(j)=\vecx(i+j-1)$ and 
    $\vecz_i(j+1)=\vecx(i+j)$. 
    Since $i+j\leq i+k-i=k$, we have $\vecx(i+j-1)>\vecx(i+j)$ 
    and hence $\vecz_i(j)>\vecz_i(j+1)$. 
    Thus, $k(\vecz_i)\geq k-i+1$. 
    Note $\vecz_i(k-i+1)=\vecx(i+(k-i+1)-1)=\vecx(k)$ 
    and $\vecz_i(k-i+2)=\vecx(i+(k-i+2)-1)=\vecx(k+1)$. 
    By definition, $\vecx(k)\leq\vecx(k+1)$ and hence 
    $\vecz_i(k-i+1)\leq\vecz_i(k-i+2)$ 
    Therefore, $k(\vecz_i)=k-i+1$. 
    As we have already seen, $\vecz_i(k(\vecz_i)+1)
    =\vecz_i(k-i+2)=\vecx(k+1)$. 
  \end{proofofclaim}

  \begin{claim}
    \label{claim:6}
    For every $i\in\{1, \ldots, l-1\}$, 
    if $\vecx(i)>\vecx(i+1)+1$, then 
    $l(\vecz_i)=l-i+1$ and hence $\vecz_i(l(\vecz_i)+2)=\vecx(l+2)$. 
  \end{claim}

  \begin{proofofclaim}
    


    Since $\vecx(i)>\vecx(i+1)+1$, 
    we have $\vecz_i(1)>\vecz_i(2)$. 
    Since $i<l$ and $\vecx(i)>\vecx(i+1)+1$, 
    we have $\vecz_i(2)=\vecx(i+1)>\vecx(i+2)+1=\vecz_i(3)+1$. 
    Thus, $l(\vecz_i)\geq 2$. 

    Let $j\in\{2, \ldots, l-i\}$. 
    Then, $\vecz_i(j)=\vecx(i+j-1)$, 
    $\vecz_i(j+1)=\vecx(i+j)$, 
    and $\vecz_i(j+2)=\vecx(i+j+1)$. 
    Note $i+j-1\leq i+(l-i)-1=l-1$ 
    and hence $i+j\leq l<k$. 
    So, either $\vecx(i+j-1)=\vecx(i+j)+1$ 
    or $\vecx(i+j)>\vecx(i+j+1)+1$. 
    Thus, either $\vecz_i(j)=\vecz_i(j+1)+1$ 
    or $\vecz_i(j+1)>\vecz_i(j+2)+1$. 
    Hence $l(\vecz_i)\geq l-i+1$. 
    If $l=k-1$, then 
    by Claim \ref{claim:5}, $k(\vecz_i)=k-i+1$ 
    and hence $l(\vecz_i)\leq k(\vecz_i)-1=k-i=l-i+1$. 
    Thus, $l(\vecz_i)=l-i+1$. 
    If $l<k-1$, then 
    we have both $\vecx(l)>\vecx(l+1)+1$ and 
    $\vecx(l+1)=\vecx(l+2)+1$. 
    Note $l-i+1\geq l-(l-1)+1=2$. 
    So, $\vecz_i(l-i+1)=\vecx(l)$, 
    $\vecz_i(l-i+2)=\vecx(l+1)$ and 
    $\vecz_i(l-i+3)=\vecx(l+2)$. 
    Thus, $\vecz_i(l-i+1)>\vecz_i(l-i+2)+1$ 
    and $\vecz_i(l-i+2)=\vecz_i(l-i+3)+1$. 
    So, $l(\vecz_i)=l-i+1$. 
  \end{proofofclaim}

  \begin{claim}
    \label{claim:7}
    For every $i\in\{1, \ldots, l-1\}$, 
    if $\vecx(i)>\vecx(i+1)+1$, 
    then $f(\vecz_i)=f(\vecx)$. 
  \end{claim}

  \begin{proofofclaim}
    By Claim \ref{claim:5}, $\vecz_i(k(\vecz_i)+1)=\vecx(k+1)$. 
    By Claim \ref{claim:6}, $\vecz_i(l(\vecz_i)+2)=\vecx(l+2)$. 
    Therefore, 
    \begin{align*}
      f(\vecz_i)&=\max\{\vecz_i(l(\vecz_i)+2), \vecz_i(k(\vecz_i)+1)\}\\
      &=\max\{\vecx(l+2), \vecx(k+1)\}=f(\vecx)
    \end{align*}
  \end{proofofclaim}

  \setcounter{case}{0}
  \begin{case}
    $m+2\leq l$. 
  \end{case}

  Then, by Claim \ref{claim:7}, 
  $\vecy(m)=\vecy(m+1)=f(\vecx)$. 
  By Claim \ref{claim:4}, 
  $k(\vecy)\geq m-1$ and $l(\vecy)\geq m-2$. 
  By Lemma \ref{lem:D}, 
  $f(\vecy)=f(\vecx)$. 

  \begin{case}
    $m+1=l$. 
  \end{case}

  Then by Claim \ref{claim:7}, $\vecy(m)=f(\vecx)$. 

  \begin{subcase}
    $l+1=k$. 
  \end{subcase}
  
  Then, we have $f(\vecx)=\vecx(k+1)$ and 
  \begin{align*}
    \vecy(m+2)&=\vecy(l+1)=\vecy(k)=\vecx(k+1)=f(\vecx)
  \end{align*}
  Consider $\vecz_l$. 
  We have 
  \begin{align*}
    \vecz_l(1)&=\vecx(l)-1\\
    \vecz_l(2)&=\vecx(l+1)=\vecx(k)\\
    \vecz_l(3)&=\vecx(l+2)=\vecx(k+1)
  \end{align*}
  By the definition of $k$, $\vecx(k)\leq\vecx(k+1)$. 
  Thus, $\vecz_l(2)\leq\vecz_l(3)$. 
  By Lemma \ref{lem:C}, 
  $f(\vecz_l)$ is either $\vecz_l(2)$ or $\vecz_l(3)$, 
  i.e. either $\vecx(k)$ or $\vecx(k+1)$. 
  In particular, we have 
  $f(\vecz_l)\leq\vecx(k+1)$ and hence 
  $\vecy(m+1)=\vecy(m)\leq\vecx(k+1)$. 
  Therefore, we have $k(\vecy)\geq m-1$, 
  $l(\vecy)\geq m-2$, $\vecy(m)=\vecy(m+2)=\vecx(k+1)$, 
  and $\vecy(m+1)\leq\vecy(m)$. 
  By Lemma \ref{lem:F}, we have $f(\vecy)=\vecx(k+1)=f(\vecx)$. 
  
  \begin{subcase}
    $l<k-1$. 
  \end{subcase}
  Then, we have $\vecx(l)>\vecx(l+1)+1$. 
  By Claim \ref{claim:5}, 
  $k(\vecz_l)=k-l+1$ and $\vecz_l(k(\vecz_l)+1)=\vecx(k+1)$. 

  \begin{subsubcase}
    $\vecx(l+2)\leq\vecx(k+1)$
  \end{subsubcase}

  Then, $\vecz_l(3)=\vecx(l+3-1)=\vecx(l+2)\leq\vecx(k+1)$
  By Lemma \ref{lem:E}, 
  $f(\vecz_l)=\vecz_l(k(\vecz_l)+1)=\vecx(k+1)$. 
  Therefore, we have $\vecy(m)=\vecy(m+1)=\vecx(k+1)$. 
  Since $k(\vecy)\geq m-1$ and $l(\vecy)\geq m-2$, 
  by Lemma \ref{lem:D}, 
  we have $f(\vecy)=\vecy(m)=\vecx(k+1)=f(\vecx)$. 

  \begin{subsubcase}
    $\vecx(l+2)>\vecx(k+2)$
  \end{subsubcase}

  Recall $k(\vecz_l)=k-l+1$. 
  Note $k-l+1\geq 1+1=2$. 
  By Lemma \ref{lem:G}, 
  \begin{align*}
    f(\vecz_l)&\leq\max\{\vecz_l(3), \vecz_l(k(\vecz)+1)\}\\
    &=\max\{\vecx(l+2), \vecx(k+2)\}=\vecx(l+2)
  \end{align*}
  Therefore, $\vecy(m+1)=\vecy(l)\leq\vecx(l+2)$. 
  Since $l<k-1$, we have $\vecx(l+1)=\vecx(l+2)+1$. 
  So, $\vecz_{l+1}(1)=\vecx(l+1)-1=\vecx(l+2)=\vecz_{l+1}(2)$. 
  Therefore, $f(\vecz_{l+1})=\vecz_{l+1}(2)=\vecx(l+2)$. 
  Hence, 
  \begin{align*}
    \vecy(m)&=\vecx(l+2)\\
    \vecy(m+1)&\leq\vecx(l+2)\\
    \vecy(m+2)&=\vecx(l+2)
  \end{align*}
  By Lemma \ref{lem:F}, 
  we have $f(\vecy)=\vecy(m)=\vecx(l+2)=f(\vecx)$. 

  \begin{case}
    $m=l$
  \end{case}

  \begin{subcase}
    $l=k-1$. 
  \end{subcase}

  \begin{claim}
    \label{claim:8}
    $\vecy(l)=f(\vecz_l)$ is either $\vecx(k)$ or $\vecx(k+1)$. 
    In particular, $\vecy(l)=f(\vecz_l)\leq\vecx(k+1)=f(\vecx)$. 
  \end{claim}

  \begin{proofofclaim}
    Since we assumed $l=k-1$, 
    $\vecy(l+1)=\vecy(k)=\vecx(k+1)$. 
    Note
    \begin{align*}
      \vecz_l(2)&=\vecx(l+1)=\vecx(k)\\
      &\leq\vecx(k+1)=\vecx(l+2)=\vecz_l(3)
    \end{align*}
    Thus, by Lemma \ref{lem:C}, 
    $f(\vecz_l)$ is either $\vecx(k)$ or $\vecx(k+1)$. 
  \end{proofofclaim}

  \begin{claim}
    \label{claim:9}
    $k(\vecy)\leq l$. 
  \end{claim}

  \begin{proofofclaim}
    Because 
    \begin{align*}
      \vecy(l)&\leq\vecx(k+1)=\vecy(k)=\vecy(l+1)
    \end{align*}
  \end{proofofclaim}

  \begin{subsubcase}
    $l=1$
  \end{subsubcase}
  Then, by Claim \ref{claim:8}, 
  $\vecy(1)=\vecy(l)\leq\vecx(k+1)$. 
  By Claim \ref{claim:1}, $\vecy(2)=\vecx(k)=\vecx(k+1)$. 
  So, $f(\vecy)=\vecx(k+1)=f(\vecx)$. 

  \begin{subsubcase}
    $l\geq 2$ and $\vecx(l)\leq\vecx(k+1)$. 
  \end{subsubcase}
  Recall that by Claim \ref{claim:8}, $\vecy(l)$ is 
  either $\vecx(k+1)$ or $\vecx(k)$. 
  If $\vecy(l)=\vecx(k+1)$, then 
  since $\vecy(l-1)=\vecx(l)\leq\vecx(k+1)=\vecy(l)$, 
  we have $k(\vecy)\leq l-1$. 
  Since we also know $k(\vecy)\geq m-1=l-1$ by Claim \ref{claim:4}, 
  we have $k(\vecy)=l-1$. 
  Since $l-2=m-2\leq l(\vecy)\leq k(\vecy)-1=l-2$, 
  we have $l(\vecy)=l-2$. 
  Therefore, $f(\vecy)=\vecx(k+1)=f(\vecx)$. 

  Suppose $\vecy(l)=\vecx(k)$. 
  Since $l<k$, we have $\vecy(l-1)=\vecx(l)>\vecx(k)=\vecy(l)$. 
  Thus, $k(\vecy)\geq l$.
  By Claim \ref{claim:9}, $k(\vecy)\leq l$ and hence $k(\vecy)=l$. 
  Note that $l(\vecy)\geq m-2=l-2=k(\vecy)-2$. 
  By Lemma \ref{lem:B}, 
  $f(\vecy)=\vecy(k(\vecy)+1)=\vecy(l+1)=\vecx(k+1)=f(\vecx)$. 
  Therefore, in either case, we get $f(\vecy)=f(\vecx)$. 

  \begin{subsubcase}
    $l\geq 2$ and $\vecx(l)>\vecx(k+1)$. 
  \end{subsubcase}
  We have $\vecy(l-1)=\vecx(l)>\vecx(k+1)=\vecy(l)$. 
  Hence, $k(\vecy)\geq l$. 
  By Claim \ref{claim:9}, we have $k(\vecy)=l$. 
  Note that $l(\vecy)\geq m-2=l-2=k(\vecy)-2$. 
  By Lemma \ref{lem:B}, 
  we have $f(\vecy)=\vecy(k(\vecy)+1)=\vecy(l+1)=\vecx(k+1)=f(\vecx)$. 

  \begin{subcase}
    $l<k-1$. 
  \end{subcase}

  \begin{claim}
    \label{new_claim:10}
    $\vecy(l+1)=\vecx(l+2)$
  \end{claim}

  \begin{proofofclaim}
  Since $l<k-1$, 
  we have $\vecx(l)>\vecx(l+1)+1$ and $\vecx(l+1)=\vecx(l+2)+1$. 
  Thus, $\vecy(l+1)=\vecx(l+2)$. 
  \end{proofofclaim}

  \begin{subsubcase}
    $\vecx(l+2)\geq\vecx(k+1)$
  \end{subsubcase}
  
  By Claim \ref{claim:5}, 
  $k(\vecz_l)=k-l+1$ and $\vecz_l(k(\vecz_l)+1)=\vecx(k+1)$. 
  We have
  \begin{align*}
    l&<k-1\\
    1&<k-l\\
    2&<k-l+1
  \end{align*}
  By Lemma \ref{lem:G}, 
  \begin{align*}
    \vecy(l)=f(\vecz_l)&\leq\max\{\vecz_l(3), \vecz_l(k(\vecz_l)+1)\}\\
    &=\max\{\vecx(l+2), \vecx(k+1)\}=\vecx(l+2)=f(\vecx)
  \end{align*}

  Then, $\vecy(l)\leq\vecx(l+2)=\vecy(l+1)$. 
  So, we have $k(\vecy)\leq l$. 
  If $l=1$, then clearly $k(\vecy)=1=l$. 
  If $l\geq 2$, then since $l<l+2\leq k$, 
  \begin{align*}
    \vecy(l-1)&=\vecx(l)>\vecx(l+2)\geq\vecy(l)
  \end{align*}
  So, $k(\vecy)\geq l$ and hence $k(\vecy)=l$. 
  Therefore, in either case, we get $k(\vecy)=l$. 

  We also have $l(\vecy)\geq m-2=l-2=k(\vecy)-2$. 
  By Lemma \ref{lem:B}, 
  $f(\vecy)=\vecy(k(\vecy)+1)=\vecy(l+1)=\vecx(l+2)=f(\vecx)$. 
  \bigskip

  Now we concentrate on the case $\vecx(l+2)<\vecx(k+1)$. 

  \begin{claim}
    \label{claim:12}
    If $\vecx(l+2)<\vecx(k+1)$, then 
    $\vecy(l)=\vecx(k+1)=f(\vecx)$. 
  \end{claim}

  \begin{proofofclaim}
    Since $l<k-1$, we have $\vecx(l)>\vecx(l+1)+1$. 
    By Claim \ref{claim:5}, 
    $\vecz_l(k(\vecz_l)+1)=\vecx(k+1)$. 
    In addition, $\vecz_l(3)=\vecx(l+2)<\vecx(k+1)$. 
    By Lemma \ref{lem:E}, 
    $\vecy(l)=f(\vecz_l)=\vecx(k+1)$.     
  \end{proofofclaim}

  \begin{subsubcase}
    $\vecx(l+2)<\vecx(k+1)$ and $k(\vecy)=l-1$. 
  \end{subsubcase}

  Note $l(\vecy)\geq l-2\geq k(\vecy)-1$. 
  By Lemma \ref{lem:A}, 
  $f(\vecy)=\vecy(k(\vecy)+1)=\vecy(l)=f(\vecx)$. 

  \begin{subsubcase}
    $\vecx(l+2)<\vecx(k+1)$ and $k(\vecy)\geq l$. 
  \end{subsubcase}

  Let $p$ be the least such that 
  $l+1\leq p<k$ 
  and $\vecx(p)>\vecx(p+1)+1$ if exists. 
  Otherwise, let $p=k$. 

  \begin{claim}
    \label{claim:13}
    $p\geq l+2$. 
  \end{claim}

  \begin{proofofclaim}
    By definition, $p\geq l+1$. 
    Since $l<k-1$, 
    $\vecx(l+1)=\vecx(l+2)+1$. 
    So, $p\geq l+2$. 
  \end{proofofclaim}

  \begin{claim}
    \label{claim:14}
    $k(\vecy)=p-1$. 
  \end{claim}

  \begin{proofofclaim}
    By assumption, we have $k(\vecy)\geq l$. 

    \begin{subclaim}
      $\vecy(l)>\vecy(l+1)$. 
      In particular, $k(\vecy)\geq l+1$. 
    \end{subclaim}
    
    \begin{proofofsubclaim}
      By Claim \ref{claim:12}, 
      $\vecy(l)=\vecx(k+1)$. 
      By assumption, $\vecx(k+1)>\vecx(l+2)=\vecy(l+1)$. 
    \end{proofofsubclaim}

    \begin{subclaim}
      \label{subclaim:14c}
      For every $i\in\{l+1, \ldots, p-2\}$, 
      $\vecy(i)=\vecy(i+1)+1$. 
      In particular, $k(\vecy)\geq p-1$. 
    \end{subclaim}

    \begin{proofofsubclaim}
      By the definition of $p$, 
      since $l<i<i+1<p$, 
      $\vecx(i)=\vecx(i+1)+1$ and $\vecx(i+1)=\vecx(i+2)+1$. 
      Thus, $\vecy(i)=\vecx(i+1)$ and $\vecy(i+1)=\vecx(i+2)$. 
      So, 
      $\vecy(i)=\vecx(i+1)=\vecx(i+2)+1=\vecy(i+1)+1$. 
    \end{proofofsubclaim}

    \begin{subclaim}
      $\vecy(p)=\vecx(k+1)$. 
    \end{subclaim}

    \begin{proofofsubclaim}
      If $p=k$, then we have $\vecy(p)=\vecy(k)=\vecx(k+1)$. 
      If $p=k-1$, then we have 
      $\vecz_p(1)=\vecx(p)-1>\vecx(p+1)=\vecz_p(2)$ 
      and $\vecz_p(2)=\vecx(p+1)=\vecx(k)\leq\vecx(k+1)
      =\vecz_p(3)$. 
      So, $\vecy(p)=f(\vecz_p)=\vecx(k+1)$. 

      Suppose $p\leq k-2$. 
      By Claim \ref{claim:5}, 
      $\vecz_p(k(\vecz_p)+1)=\vecx(k+1)$. 
      Note $l+2\leq p\leq p+2\leq k$, so, 
      $\vecz_p(3)=\vecx(p+2)\leq\vecx(l+2)<\vecx(k+1)=\vecz_p(k(\vecz_p)+1)$. 
      By Lemma \ref{lem:E}, 
      $y(p)=f(\vecz_p)=\vecz_p(k(\vecz_p)+1)=\vecx(k+1)$. 
    \end{proofofsubclaim}

    \begin{subclaim}
      \label{subclaim:14e}
      $\vecy(p-1)<\vecx(k+1)=\vecy(p)$. 
      In particular, $k(\vecz)\leq p-1$. 
    \end{subclaim}

    \begin{proofofsubclaim}
      Since $p\geq l+2$, we have $p-1\geq l+1>l$. 
      By the definition of $p$, 
      we have $\vecx(p-1)=\vecx(p)+1$ and hence $\vecy(p-1)=\vecx(p)$. 
      Since $l+2\leq p\leq k$, 
      we have $\vecx(p)\leq\vecx(l+2)$. 
      By assumption, $\vecx(l+2)<\vecx(k+1)$. 
      Therefore, $\vecy(p-1)<\vecx(k+1)$. 
    \end{proofofsubclaim}

    By Subclaim \ref{subclaim:14c} and Subclaim \ref{subclaim:14e}, 
    we have $k(\vecy)=p-1$. 
  \end{proofofclaim}

  If $l(\vecy)=k(\vecy)-1$, then clearly 
  $f(\vecy)=\vecy(k(\vecy)+1)=\vecy(p)=\vecx(k+1)=f(\vecx)$. 
  Suppose $l(\vecy)<k(\vecy)-1$. 
  By Claim \ref{claim:4}, 
  $l(\vecy)\geq m-1=l-1$. 
  Thus, $l+1\leq l(\vecy)+2$. 
  Recall $\vecy(l+1)=\vecx(l+2)<\vecx(k+1)=\vecy(p)=\vecy(k(\vecy)+1)$. 
  By Lemma \ref{lem:A}, 
  $f(\vecy)=\vecy(k(\vecy)+1)=f(\vecx)$. 
\end{proofoflem}

\begin{lem}
  For every $\vec{x}\in\Z^n$, $f(\vec{x})$ satisfies the tarai recurrence. 
\end{lem}

\begin{proofoflem}
  By Lemma \ref{k(x)_lt_n}, we may assume that $k(\vec{x})=n$, i.e. 
  $\vec{x}(1)>\vec{x}(2)>\cdots>\vecx(n)$. 
  For every $i=1,\cdots, n$, define 
  $\vecz_i=\sigma(r^{i-1}(\vecx))$ and 
  $\vecy(i)=f(\vecz_i)$. 
  We need to show that $f(\vecy)=f(\vecx)=\vecx(1)$. 

  \begin{claim}
    $\vecy(1)<\vecx(1)$. 
  \end{claim}
  
  \begin{proofofclaim}
    We have $\vecz_1(1)=\vecx(1)-1$ and $\vecz_1(i)=\vecx(i)$ 
    for every $i=2, \ldots, n$. 
    Then, clearly we have $\vecy(1)=f(\vecz_1)\leq\max\vecz_1=\vecx(1)-1$. 
  \end{proofofclaim}

  \begin{claim}
    $\vecy(n)=\vecx(1)$
  \end{claim}

  \begin{proofofclaim}
    Since $\vec{z}_n(1)=\vec{x}(n)$ and $\vec{z}_n(2)=\vecx(1)$, 
    we have
    $\vecy(n)=f(\vecz_n)=\vecz(2)=\vecx(1)$. 
  \end{proofofclaim}

  \begin{claim}
    \label{claim:fin_3}
    For every $i=2, \ldots,n-1$, 
    either $\vecy(i)=\vecx(1)$ or $\vecy(i)=\vecx(i+1)$. 
  \end{claim}

  \begin{proofofclaim}
    Note 
    \begin{align*}
      \vecz_i(1)&=\vecx(i)-1\\
      \vecz_i(j)&=\vecx(j+i-1)\text{ (for all $j=2, \ldots, n-i+1$)}\\
      \vecz_i(n-i+2)&=\vecx(1)
    \end{align*}

    If $\vecx(i)-1=\vecx(i+1)$, then 
    we have $\vecz_i(1)=\vecz_i(2)$ and hence 
    $\vecy(i)=f(\vecz_i)=\vecz_i(2)=\vecx(i+1)$. 

    Suppose $\vecx(i)-1>\vecx(i+1)$. 
    Then, for every $j=2, \ldots, n-i$, we have
    $\vecz_i(j)=\vecx(j+i-1)>\vecx(j+i)=\vecz_i(j+1)$. 
    Moreover, $\vecz_i(n-i+1)=\vecx(n)<\vecx(1)=\vecz_i(n-i+2)$. 
    So, $k(\vecz_i)=n-i+1$ and 
    $\vecz_i(k(\vecz_i)+1)=\vecz_i(n-i+2)=\vecx(1)$. 
    Then, since $\vecz_i(2)=\vecx(i+1)<\vecx(1)=\vecz_i(k(\vecz_i)+1)$, 
    by Lemma \ref{lem:H}, 
    we have $\vecy(i)=f(\vecz_i)=\vecz_i(k(\vecz_i)+1)=\vecx(1)$. 
  \end{proofofclaim}

  Let $m$ be the least such that $\vecy(m+1)=\vecx(1)$. 
  Since $\vecy(n)=\vecx(1)$, there is such an $m\leq n-1$. 
  If $m=1$, then we have $\vecy(1)<\vecx(1)=\vecy(2)$ and hence 
  $f(\vecy)=\vecy(2)=\vecx(1)$. 

  Suppose $m>1$. 
  Then for every $i=2, \ldots, m$, by Claim \ref{claim:fin_3}, we have
  $\vecy(i)=\vecx(i+1)$. 
  Hence, for every $i=2, \ldots, m-1$, we have
  $\vecy(i)>\vecy(i+1)$. 
  We also have $\vecy(m)=\vecx(m+1)<\vecx(1)=\vecy(m+1)$. 
  Therefore, $k(\vecy)=m$ and $\vecy(m+1)=\vecx(1)$. 
  In particular, $\vecy(2)=\vecx(3)<\vecx(1)=\vecy(m+1)=\vecy(k(\vecy)+1)$. 
  By Lemma \ref{lem:H}, 
  we have $f(\vecy)=\vecy(k(\vecy)+1)=\vecx(1)$. 
\end{proofoflem}
\def\cprime{$'$}

\end{document}